\documentstyle[12pt,fullpage,epsf]{article}

\begin{document}

\def\Cttw{$C(t+t_{\rm w}, t_{\rm w})$}
\def\Ebt{$E_{\rm B}(t)$}
\def\mCttw{$m_{\rm C}(t; t_{\rm w})$}
\def\mttw{$m(t; t_{\rm w})$}
\def\nle{\ \raise.3ex\hbox{$<$}\kern-0.8em\lower.7ex\hbox{$\sim$}\ }
\def\nge{\ \raise.3ex\hbox{$>$}\kern-0.8em\lower.7ex\hbox{$\sim$}\ }
\def\Pqt{$P(q; t)$}
\def\qbar{${\bar q}$}
\def\qc{$q_{\rm c}$}
\def\qEA{$q_{\rm EA}$}
\def\Qttw{$Q(t+t_{\rm w}, t_{\rm w})$}
\def\terg{$t_{\rm erg}(N, T)$}
\def\tergE{$t_{\rm erg}^{\rm E}(N, T)$}
\def\tergS{$t_{\rm erg}^{\rm S}(N, T)$}
\def\tergL{$t_{\rm erg}^{\rm L}(N, T)$}
\def\tergMY{$t_{\rm erg}^{\rm MY}(N, T)$}
\def\tergtr{$t_{\rm erg}^{\rm tr}(N, T)$}
\def\tcrs{$t_{\rm crs}$}
\def\ta{$t_{\rm a}$}
\def\Tc{$T_{\rm c}$}
\def\tmax{$t_{\rm max}$}
\def\tmin{$t_{\rm min}$}
\def\tqe{$t_{\rm q.e.}$}
\def\tw{$t_{\rm w}$}
\def\twone{$t_{{\rm w}1}$}

\vspace*{2cm}
\noindent
\begin{center}
{\bf Monte Carlo Simulation on Aging Processes
within\\ One `Pure State' of the SK Spin-Glass Model}

\vspace*{1cm}
Hajime Takayama\footnote{
takayama@issp.u-tokyo.ac.jp}, Hajime Yoshino
\footnote{yhajime@ginnan.issp.u-tokyo.ac.jp} 
and Koji Hukushima \footnote{fukusima@ginnan.issp.u-tokyo.ac.jp}

\vspace*{1cm}
Institute for Solid State Physics, University of Tokyo,\\
7-22-1  Roppongi, Minato-ku, Tokyo 106, Japan

\vspace*{1cm}

\vspace*{1cm}
submitted to Journal of Physics. A: Mathematical and General

\end{center}


\vspace*{1cm}
\begin{abstract}
Monte Carlo simulations on the SK model have been done to 
investigate aging processes after a rapid quench 
from $T=\infty$ to the spin-glass phase. 
The time range of simulations is taken care of 
so that the system does not surmount free-energy barriers 
between the pure states which are considered to diverge 
in the thermodynamic limit. 
We simulate time evolutions of extensive energy of the system, 
Parisi's overlap distribution function, auto-correlation 
and clones-correlation functions, 
distribution functions of the two correlations, and magnetization 
induced by the field applied after a certain waiting time. 
The data simulated exhibit a rich variety of aging phenomena. 
Most of them can be interpreted in a 
unified way, though qualitatively, by the scenario of 
{\it growth of quasi-equilibrium 
domains} which we will introduce in this work. 
More surprisingly, they suggest strongly that a basin of 
attraction of one dominant pure state spans 
almost an entire phase space of the system with a common 
time-reversal symmetry. 
\end{abstract}

\clearpage
\vspace*{-3mm} 
\section{Introduction}

Since the first observation by Lundgren et al\cite{Lundg} 
aging phenomena in spin glasses have been extensively 
studied\cite{exp-review}.
They are expected to reveal nature of the low-temperature 
spin-glass phase which has not yet been settled in spite of nearly 
two decades of dispute. One of the key concepts on aging phenomena 
recently introduced by Bouchaud\cite{Bouch} 
and incorporated in 
the analytic theories\cite{analytic}\cite{analy-2} 
is {\it weak-ergodicity 
breaking}. It was originally introduced in the argument on the 
trap model\cite{Bouch}. In a system exhibiting aging phenomena 
there exist many metastable states, or 'traps'. 
Depths of the traps are distributed continuously in such a way 
that the average time for the system to escape from them by thermal 
activation process is infinite; or, it takes an 
infinite time for the system to equilibriate. 
This picture of weak-ergodicity breaking is represented 
in terms of the spin auto-correlation function 
$C(t,t') \equiv N^{-1}\sum_i S_i(t)S_i(t')$ as 
follows\cite{Bouch}\cite{analy-2}:
$$ {\partial C(t,t') \over \partial t} \le 0, 
\ \ \ {\rm and}\ \ \ 
{\partial C(t,t') \over \partial t'} \ge 0,
\eqno{(1a)} $$ 
$$ \lim_{t\rightarrow\infty}\lim_{t'\rightarrow\infty}
C(t,t') = q_{\rm c} \ \ (>0), 
\eqno{(1b)} $$
$$ \lim_{t\rightarrow\infty}
C(t,t') = 0  \ \ \ \ \ {\rm for \ a \ fixed} \ \ t' \ \ (< \infty), 
\eqno{(1c)} $$
where \qc\ is a certain constant. 

Various simulations which reproduce some aspects of the 
experimentally observed 
aging phenomena have been already 
reported\cite{AMS}\cite{CK-simu}\cite{Rieger}\cite{F-Rieger}.
But they are not yet satisfactory to be able to 
understand the phenomena in a unified way. 
Since the recent phenomenological and theoretical arguments mentioned 
above, as well as recent analyses on 
experiments\cite{VHOBC}\cite{Joh}, 
are more or less based on the mean-field picture of spin 
glasses, we believe it is worth examining aging phenomena 
in the mean-field spin-glass model more in detail 
by simulations\cite{Bal}.  
In the present work therefore we have performed Monte Carlo (MC) 
simulation on the SK model\cite{SK}, whose equilibrium properties 
are well understood\cite{MPV}. Its purpose is to obtain 
various information on its aging phenomena, 
including those which are related to the above-mentioned key 
concept.

Aging phenomena we are concerned with are non-equilibrium 
processes toward equilibrium observed in macroscopic systems in 
long but finite intervals of time. Since, on the other hand, the 
present simulations are performed 
only on finite systems with $N$ spins, we 
have to take care of their time range.  Suppose the system has a 
certain relaxation time, \terg, which may depend also on temperature 
$T$, and whose systematic dependence on $N$ is known. If 
lim$_{N\rightarrow\infty}$\terg\ $=\infty$, simulations on aging 
phenomena have to be done in the time range of $t\nle$\terg\ in order 
to be free from the finite-size effect.

According to Parisi's picture\cite{MPV} on the spin-glass phase of 
the SK model in equilibrium, there are many pure states 
which are separated from each other by insurmountable free-energy 
barriers. More explicitly, it was 
argued that in systems with large $N$ there exist two 
types of the free-energy barriers: one between the pure states 
with a common time-reversal symmetry, and the other between those 
with opposite time-reversal symmetries\cite{MacYoun}\cite{BhattY}.
The characteristic times, \tergS\ and \tergL, needed to surmount 
respectively the former and latter barriers are given by 
(see Fig. 2 below)
$$  {\rm ln} \ t_{\rm erg}^{\rm S}(N, T) \propto N^{1/4},
\ \ \ {\rm and} \ \ \ 
{\rm ln} \ t_{\rm erg}^{\rm L}(N, T) \propto N^{1/2}.
\eqno{(2)} $$ 
Therefore we have to set at least \tmax$<$\tergS, 
where \tmax\ is the maximum observation time of simulations.
In the present work we call such a limited 
time interval after a rapid quench  
{\it aging range}, and a portion of phase space that the system 
explores within that interval  a {\it `pure state'}. 

In the present work we simulate time evolution of various quantities 
after the quench; extensive energy of the system, Parisi's overlap 
distribution function, auto-correlation and  
clones-correlation functions\cite{Bal}\cite{Clone-2}\cite{Clone-3}, 
and distribution functions of the two correlations. 
The magnetization induced by the field applied after a waiting time 
\tw\ is also investigated. 
It has turned out that the data simulated exhibit 
a rich variety of aging phenomena. They indicate complexity 
of the rugged free-energy structure within the `pure state'. 
It is expected to have many free-energy barriers of finite heights 
which are surmountable within the aging range introduced 
above. Most of our simulated results can be interpreted in a 
unified way, though qualitatively, 
by the scenario of {\it growth of quasi-equilibrium 
domains} which we will introduce and explain in this work. 
Surprisingly, the results suggest that the basin of 
attraction of one `pure state' spans 
an almost entire phase space with a common 
time-reversal symmetry. We call this feature the 
{\it dynamic dominant pure-states picture}. 
On equilibrium properties of the SK model, on the other hand, 
there is the argument\cite{MPSTV} that 
among many pure states 
only a few (even one) of them dominate the Gibbs-Boltzmann 
measure. We call this aspect
the {\it static dominant pure-states picture}. 
Our simulated data imply that the dynamic dominant pure-states 
picture describes dynamical (or aging) processes in the phase 
space whose organization of states gives rise to the 
static dominant pure-states picture for equilibrium.
 
In the next section the model and method of our simulations 
are briefly explained.
In section 3 we present the results of simulations which are 
interpreted by our scenario of growth of quasi-equilibrium 
domains in section 4. The final section is devoted to the concluding 
remarks. 

\vspace*{-3mm} 
\section{Model and Method}

By means of the standard heat-bath method of the MC 
simulation we have studied the 
$\pm J$ SK model with mean zero and variance 
$(N-1)^{-1/2}$\cite{pmJSK}. 
The system sizes examined are $N=32\sim2048$, particularly 
$N=128, \ 512$ in detail.
Mainly analyzed in the present work are aging processes after 
rapid quenches from $T=\infty$ to $T$ below the spin-glass transition
temperature \Tc\ ($=1$ in the limit 
$N\rightarrow\infty$). 
Simulation of this process is to simply perform a MC run at $T$ 
starting from a random spin configuration. 
Physical quantities are obtained by taking average over such $M$ 
independent MC runs 
for each realizations of $\{ J_{ij} \}$ (sample) and over the 
$N_{\rm s}$ samples.  Those at time $t$ (in unit of 1MC step 
per spin) in each MC run are evaluated 
by means of the following short time average,  
$$ A^{(m)}(t) = {1 \over \Delta t + 1} \sum_{t'=t-\Delta t}^t 
a^{(m)}(t'),
\eqno{(3)} $$
where $a^{(m)}(t')$ is the value of quantity $A$ at step $t'$ of 
$m$-th MC run, and 
$\Delta t \ll t$ (eg. $\Delta t=0$ for $t<16$ and 511 for $t \ge 
2^{13}$). In the present work we put $M=10$ and 
$N_{\rm s}=40\sim2000$ depending on $N$. 

Most of our simulations are performed in the aging range 
mentioned in the previous section.
Here we further make a comment on this time range in relation to the
limiting procedures in eqs.(1b, c). For these equations the limit 
$N\rightarrow\infty$ is taken before those of $t'$ and/or $t$. 
Such limiting procedures cannot be carried out in our simulational
study on finite systems. In the present work limiting values of the
types of eqs.(1b) and (1c) are extrapolated from the following
quantities in systems with finite $N$: those simulated in the time
range  $t\ll t'$ for the type of eq.(1b) and thoes in $t\gg t'$ for 
that of eq.(1c), both $t$ and $t'$ being in the aging range.
   
Another comment is on a cut-off of the aging range in the shorter
side. The time range of simulations should not be too short 
in order to distinguish 
aging phenomena peculiar to spin glasses from certain 
initial transient processes expected to exist in any systems. 
We denote a time-scale of this short cut-off as \tqe. 
Its explicit value, as well as that of \tmax, will be discussed 
below referring to our simulated data. 

\vspace*{-3mm} 
\section{Results}

\vspace*{-3mm} 
\subsection{Energy and Parisi's overlap distribution}

Let us first investigate the time evolution of the energy 
$E(t_{\rm a})$ after the quench at \ta$=0$. 
As shown in Fig. 1a, the energy per spin drops rapidly 
and then gradually 
saturates to the $N$-dependent equilibrium value $E_{\rm eq}/N$. 
Interesting information is obtained when we look at how 
the extensive energy $E(t_{\rm a})$ approaches to $E_{\rm eq}$. 
In  Fig. 1b we show $\Delta E(t_{\rm a}) 
\equiv E(t_{\rm a}) - E_{\rm eq}$ 
where $E_{\rm eq}$ is approximated by $E(t_{\rm max})$. 
The time \tergE, defined by $\Delta E(t_{\rm erg}^{\rm E}) \cong T$, 
is shown in Fig. 2.  It has the same $N$-dependence as that 
of \tergS\ of eq.(2) obtained previously\cite{MacYoun}, 
though the proportional constants are different. 
In the present work we adopt \tergE\ 
as the upper boundary of the aging range. 

A typical barrier energy 
\Ebt\ for the system to be able to surmount within time $t$ is 
given by \Ebt$\simeq T{\rm ln} t$, where the attempt time of 
thermal activation process is put 1 MC step.
The value $\Delta E(t_{\rm a})+E_{\rm B}(t_{\rm a})$ 
for $N=$512 is also 
shown in Fig. 1b. The data shown in the figure tell that relaxational 
processes in the aging range are 
associated with decrease of $\Delta E(t_{\rm a})$ (even of 
$\Delta E(t_{\rm a})+E_{\rm B}(t_{\rm a})$). 

In Fig. 3 the time evolution of Parisi's overlap distribution functions 
$P(q;t_{\rm a})$ is demonstrated for systems with $N=128$ at $T=0.4$. 
Here $q$ is one of $M(M-1)/2$ overlaps of magnetization configurations 
$q(t)$ $=$ $N^{-1}$ $\sum_i m_i^{(k)}(t)m_i^{(l)}(t)$
where $m_i^{(k)}(t)$ is evaluated by eq.(3) with $a=S_i$. 
In the figure the thick broken line represents $P_{\rm RSB}(q)$, i.e.,
$P(q)$ in equilibrium and in the thermodynamic limit 
which is evaluated by solving 
numerically Parisi's equation\cite{Nemoto}.
The position of its delta peaks are specified $\pm$\qEA; the 
self-overlap parameter\cite{MPV}. 
As already reported by Bhatt and Young\cite{BhattY}, 
$P(q;t_{\rm a})$ evolves from a single gaussian centered at $q=0$ 
at $t_{\rm a}=0$ to a structure with double peaks (but of 
significant widths) similar to $P_{\rm RSB}(q)$ 
at large enough $t_{\rm a}$. In the present analysis we 
determine \tergL\ as the time at which $P(q\simeq0;t_{\rm a})$ 
converges to $P_{\rm RSB}(q\simeq0)$ and show it also in Fig. 2. 
We see in Fig. 3 that at $t_{\rm a}=512 \simeq 
t_{\rm erg}^{\rm E}(N=128, T=0.4)$, 
$P(q;t_{\rm a})$ exhibits already the double peaks around 
$q=\pm$\qEA but its peak at $q\simeq0$ still remains. 

\vspace*{-3mm} 
\subsection{Auto-correlation function}

In Fig. 4a we show typical data of the auto-correlation function 
\Cttw. It is obtained  at $T=0.4$ 
of systems with $N=512$ in the aging range, i.e., 
$t_{\rm a}=t+t_{\rm w} \nle 
t_{\rm erg}^{\rm E}(N=512, T=0.4) \simeq 3\times10^4$. 
One can see that \Cttw\ satisfies eq.(1a), i.e., one of 
the conditions of the weak-ergodicity breaking. 
Also the data exhibit crossover behavior 
from relatively slow decay in $t <$\ \tw\ to relatively rapid 
decay in $t >$\ \tw. The behavior is common to \Cttw\ simulated 
in various systems\cite{AMS}\cite{CK-simu}\cite{Rieger}\cite{Yoshi}, 
and is interpreted as a crossover from quasi-equilibrium 
to out-of-equilibrium behaviors. 

An interesting feature 
becomes clear if we plot \Cttw\ against 
ln($t/t_{\rm w}$), as shown in Fig. 4b.
As first pointed out by Baldassarri\cite{Bal}, \Cttw\ with 
different \tw\ cross almost at a point $t\simeq$\tcrs. 
This behavior is certainly different from those of 
\Cttw\ observed in the previous 
works\cite{CK-simu}\cite{Rieger}\cite{Yoshi}, 
whose \Cttw\ with different \tw's in the out-of-equilibrium range 
are well scaled to a single curve when plotted against $t/$\tw.
We also note that 
ln\tcrs$\simeq$ln\tw\ and that the value of 
$C(t_{\rm crs}+t_{\rm w}, t_{\rm w})$ 
is rather closer to $1-T =$\qbar\ 
than \qEA\ ($\cong 0.75$), where \qbar$=\int_0^1q(x){\rm d}x$ is 
Parisi's order parameter in equilibrium\cite{MPV}. 

That the above crossing feature of \Cttw\ is a genuine property of 
aging processes in the present system is ascertained by the inspection 
of $C(t_{\rm a}, t_{\rm a}/2)$ which can be simulated by single 
aging process. As shown in Fig. 5 $C(t_{\rm a}, t_{\rm a}/2)$ for 
various $N$ and $T$ are nearly independent of $t_{\rm a}$ 
in the range $t_1\nle t_{\rm a} \nle t_2(N, T)$. 
Here $t_2(N, T)$ is a time around which $C(t_{\rm a}, t_{\rm a}/2)$ 
starts to decrease from its nearly constant value, and we have 
checked that it coincides with \tergE\ within logarithmic 
accuracy. On the other hand, the shorter time scale $t_1$ is regarded 
as \tqe\ mentioned in section 2. 
The near constancy of $C(t_{\rm a}, t_{\rm a}/2)$ implies that 
\Cttw\ with different \tw's have a nearly common value 
at ln($t/$\tw$)\simeq 0$. 

\vspace*{-3mm} 
\subsection{Clones-correlation function}

We have also investigated the clones-correlation function \Qttw\ 
\cite{Bal}\cite{Clone-2}\cite{Clone-3}. It is the correlation of 
two configurations starting from an identical one which has evolved 
up to $t_{\rm a}=$\tw, but are evolving 
by means of two independent MC 
runs (with different sets of random numbers) at $t_{\rm a}>$\tw. 
A typical result is shown in Fig. 6 for 
$T=0.4, N=512$. As found also by Baldassarri\cite{Bal}, 
\Cttw\ and \Qttw\ cross at ln$t\sim$ln\tw. 
Their values at the 
crossing are again closer to \qbar\ independently of \tw. 
A more remarkable feature seen in the figure 
is that \Qttw\ with fixed \tw\ 
tend to saturate to constant values which depend on \tw\ 
as $t\rightarrow$\ \tergE.  
It is also noted (not shown) 
that \Qttw\ plotted against ln($t/t_{\rm w}$) also cross 
nearly at a point similarly to \Cttw\ in Fig. 4b. 

\vspace*{-3mm} 
\subsection{Fluctuation-dissipation theorem}

In order to get further insights into aging processes 
we have simulated the magnetization \mttw\ 
induced by the field $h$ which is switched on at $t_{\rm a}=$\ \tw. 
The result for $T=0.4, N=512$ 
and with $h=0.1$ is drawn against ln($t/$\tw) in Fig. 7. 
Also shown by broken lines in the figure are 
\mCttw\ defined by 
$$ m_{\rm C}(t; t_{\rm w}) = h\{1-C(t+t_{\rm w}, t_{\rm w})\}/T.
\eqno{(4)} $$
The fluctuation-dissipation theorem (FDT) tells that, 
in equilibrium and for an infinitesimally small $h$, \mttw\ and 
\mCttw\ are independent of \tw\ and $m(t)=m_{\rm C}(t)$ holds. 
We see in Fig. 7 that, as has been also observed in other 
model systems\cite{AMS}\cite{F-Rieger}\cite{Yoshi}, \mttw$\cong$\mCttw\ 
holds in the time range $t \ll$\tw\ with \tw\ larger than 
a certain value. The latter value, which roughly coincides 
with $t_1$ introduced above, is also used to estimate \tqe\ which 
specifies the initial transient regime of the aging 
process.

\vspace*{-3mm} 
\section{Discussions}

\vspace*{-3mm} 
\subsection{Scenario of growth of quasi-equilibrium domains}

In order to interpret the above results of our simulations, let us 
first remind of the following properties of the SK model in 
equilibrium. It is known that the TAP equation\cite{TAP} of 
the SK model has a huge number of solutions, or local free-energy
minima, in the spin-glass phase\cite{BrayM}. 
In equilibrium, however, only a very limited 
number of the local minima have significant probability weights 
$P_\alpha (\propto {\rm exp}(-F_\alpha/T)$) where $F_\alpha$ is the 
(extensive) TAP free-energy of the $\alpha$-th 
solution\cite{MPSTV} (the {\it static dominant pure-states picture}
mentioned in \S 1). 
Associated with such lowest free-energy 
states are limited portions of phase space which are separated 
by free-energy barriers described by eq.(2). One of the possible 
interpretations is then that aging processes simulated in the present 
work reflect the free-energy structures of such local regions centered 
at each lowest free-energy state. We may identify these regions 
to the `pure states' introduced in section 1.

Based on the above assumption we argue that 
each `pure state' contains a huge number of local minima 
(which we call hereafter simply states), as schematically shown in
Fig. 8. The lower their (extensive) energies, the less is 
the number of the states. [Here and hereafter we disregard entropy 
effects which have not been evaluated explicitly 
in the present simulation.] 
Their basins of attraction for a rapid quench process are considered 
to have more or less comparable sizes\cite{NishiNT}.
Therefore after a certain initial transient time 
the system is found, with probability of almost unity, in one of the
huge number of states with a relatively higher energy, 
from which it starts to look for lower 
and lower energy states by thermal activation processes; i.e., 
it ages. We consider that \tqe\ introduced before corresponds to the
above-mentioned transient time. it is noted that \tqe\ are 
little dependent on $N$ and is less than a few tens MC steps 
(see Figs. 5 and 7).

Let us then suppose that at time $t_{\rm a}=$\ \twone\ 
in the aging range the system reaches to a certain state, say $S_1$
in Fig. 8. In a time interval of $t_{\rm a}=t_{\rm w1}+t$ with 
$t\ll$\ \twone\ the system is fluctuating in a local region $R_1$
centered at $S_1$ (the shaded area in Fig. 8), 
thereby it is expected to visit states in the 
region with frequencies according to their relative 
Boltzmann weights. We call such regions 
{\it quasi-equilibrium domains}. 
As time goes on another interval of order \twone, i.e., at 
$t_{\rm a} = t_{{\rm w}2} \simeq 2t_{{\rm w}1}$,
the system surmounts a barrier of order $T$ln\twone\ and finds a lower 
energy state, say $S_2$ in Fig. 8. This corresponds to
entering into the out-of-equilibrium range of the waiting 
time $t_{\rm w1}$.  The state $S_2$ is, in turn, associated with the  
quasi-equilibrium domain $R_2$ of the waiting 
time $t_{\rm w2}$ which is larger than $R_1$. 
The quasi-equilibrium domain grows with time until it exhausts 
the whole `pure state', or the system reaches the lowest energy 
state $S_\infty$ in Fig. 8 
(which takes an infinite time if $N=\infty$). 

The above scenario for aging processes, which we call 
{\it growth of quasi-equilibrium domains}, has been initially 
introduced from our relaxational-modes analysis 
on the same problem\cite{YHT}. From the present MC 
study it is introduced to explain 
straightforwardly the following results in the aging range 
of \tqe $\nle t_{\rm a} \nle$ \tergE: 1) the monotonic decrease of 
the extensive energy in the aging range ($\Delta E(t_{\rm a})$ in
Fig. 1b), 2) the crossover from quasi-equilibrium 
to out-of-equilibrium behaviors at around $t\sim$\tw\ seen in \Cttw\ 
of Fig. 4, 3) the apparent FDT at $t\ll$\tw\ demonstrated in Fig. 7,
and 4) the behavior of the clones-correlation shown in Fig. 6 which 
is explained as follows. 
Suppose the clones are created at \ta=\twone\ from a 
spin configuration near $S_1$ in Fig. 8. Then at 
$t\nge$\ \twone, the clones visit more frequently near the same state 
$S_2$ whose relative Boltzmann weight is largest in the 
larger domain $R_2$. The Hamming distance between them is then 
considered to be smaller than that between $S_1$ and $S_2$ which 
determines predominately the value of the corresponding 
auto-correlation function. 
This means $Q(t+t_{{\rm w}1},t_{{\rm w}1})$ $>$ 
$C(t+t_{{\rm w}1},t_{{\rm w}1})$ at 
$t \nge t_{{\rm w}1}$ as observed in the simulation. 

Concerned with the above items 2) and 3), we note that 
at $t\ll$\tw\  \Cttw\ depend not only on $t$ but also 
on \tw\ (see Fig. 4a). This means that nature of overall 
fluctuation within each quasi-equilibrium domain differs 
between the domains and from true equilibrium. Still \Cttw\ and 
\mttw\ obey the apparent FDT. This feature is certainly different 
from the recent arguments\cite{analy-2}\cite{VHOBC} that 
both \Cttw\ and \mttw\ consist of two parts, equilibrium and aging 
ones, and that the FDT holds only for the equilibrium parts 
which depend on $t$ alone. 

We also note that the above time range \tw$\gg t
(\nge$\tqe) is the one where we look for the limiting value of eq.(1b) 
as pointed out in Section 2. 
Naive speculation based on the results shown in Figs. 4b and 5
is that \qc\ in eq.(1b) is certainly less than \qEA\ in contrast to
the previous claim of \qc$=$\qEA\cite{analy-2}.
In our simulation on a large but finite
$N$ system, we regard that the system is in `equilibrium' within
one `pure state' at \tw\ of the order of \tergE. 
The extrapolation to $N=\infty$ keeping these conditions at
\tw$\sim$\tergE\ gives rise to \qc\ smaller than \qEA.

\vspace*{-3mm} 
\subsection{Distribution functions of clone- and auto-correlations}

The time evolution of distribution functions $P(Q;t,t_{\rm w})$ 
with $Q(t)$ $=$ $N^{-1}$ $\sum_i m_i^{[1]}(t)m_i^{[2]}(t)$ 
shown in Fig. 9a further supports the above interpretation of the 
clones-correlation. 
Here $m_i^{[k]}(t)$ is evaluated by eq.(3) whose $a$ in the r.h.s. 
is $S_i$ of the $k$-th clone. The distribution is therefore 
obtained from $MN_{\rm s}$ data points of $Q(t)$.  
A sharp peak of $P(Q;t,t_{\rm w})$ at $t \ll$\ \tw\ located around 
$Q \simeq$\ \qEA\ indicates simply 
that the clones are not yet separated enough and are fluctuating 
near the lowest energy state in the quasi-equilibrium domain 
of waiting time \tw. Around $t \simeq$\ \tw\ height of the peak 
decreases and its width increases. This 
is interpreted that the clones 
now tend to escape statistically independently of each other from 
the domain of \tw. Interestingly at $t \gg$\ \tw\ the peak near 
$Q \simeq$\ \qEA\ sharpens again though its height does not 
recover very much. The peak in this time range suggests 
that the clones are again near the same lowest energy state but of 
a larger quasi-equilibrium domain than the original one. 
This implies that \Qttw\ for 
fixed \tw\ converge to certain finite values. This is in fact 
the case as we have already seen in Fig. 6; \Qttw\ almost saturate to
certain constants. Correspondingly, the shape of $P(Q;t,t_{\rm w})$ 
becomes little dependent on $t$ at $t\ (\sim$\tergE).

In order to examine the $N$-dependence of the above characteristic
feature in the clones-correlation, we show in Fig. 9b \Qttw\ at
$t\simeq$\tergE\ for $N=$128 and 512 both with \tw=64. The two almost
coincide with each other. The small difference in the peak positions
near $Q=$\qEA\ is attributed to the finite-size effect. 
Naive extrapolation 
of these results to $N\rightarrow\infty$ (thereby 
\tergE\ $\rightarrow\infty$) is lim$_{t\rightarrow\infty}$\Qttw$>0$ 
for a fixed \tw\ ($<\infty$). Thus the aging process within 
one `pure state' of present interest belongs to type I by means of 
the classification introduced by Barrat et al\cite{Clone-3}.

In Fig. 9b we also show, by the thin solid curve, \Qttw\ at
$t\simeq$\tergL\ for $N=128$. One sees that a small peak starts to 
develop not around $Q=0$ but around $Q=-q_{\rm EA}$. This result is
interpreted as follow. When one of the clones surmounts the energy
barrier between the phase spaces with different time-reversal
symmetry, it goes down and situates near the counterpart of $S_\infty$ 
within a time interval of \tergE ($\ll$\tergL). 

In contrast to \Qttw\ above discussed, \Cttw\ in Fig. 6 doesn't exhibit 
tendency of the saturation to certain finite values within the aging 
range of the present concern. Although it neither completely 
vanishes in the same time range, 
we expect that the weak-ergodicity breaking 
condition of eq.(1c) holds in the limit 
$N\rightarrow\infty$ within its aging range. 
This expection is derived from inspection of the distribution functions 
of auto-correlation, $P(C;t,t_{\rm w})$ with $C(t)$ $=$ $N^{-1}$
$\sum_i S_i(t+t_{\rm w})S_i(t_{\rm w})$, shown in Fig. 10. 
At $t>t_{\rm w}$, in contrast to $P(Q;t,t_{\rm  w})$, $P(C;t,t_{\rm
  w})$ tends to become a gaussian form whose center tends to approach
$C=0$. We therefore consider that the weak-ergodicity breaking of 
eq.(1c) holds since any configuration at $t_{\rm a}=$\tw\ 
($< \infty$) in a `pure state' is orthogonal 
to (separated far away from) the configuration at 
$t_{\rm a}=\infty$, i.e., the lowest energy state of 
the `pure state'. 

According to Barrat et al\cite{Clone-3}, the above aging process
of type I judging from \Qttw\ and 
with the weak-ergodicity breaking judging from \Cttw, is expected 
to appear, for example, in a system having a `gutter' in phase space. 
But our scenario of growth of quasi-equilibrium domains is 
also compatible with the two characteristics. 
We only need a quite natural assumption that a basin of attraction 
of a `pure state' is infinitely large if $N=\infty$. In fact our 
scenario is more appropriate judging from the time evolution of 
$P(Q; t, t_{\rm w})$ already described above. We may say 
that the energy structure in one `pure state' is, 
instead of a `gutter', a `funnel' with an infinitely 
wide input-mouth.

\vspace*{-3mm} 
\subsection{Dominant pure-states pictures}

Lastly let us compare the three distribution functions 
$P(Q;t\sim t_{\rm erg}^{\rm E},t_{\rm w})$,  
$P(q; t_{\rm a}\simeq t_{\rm erg}^{\rm E})$ and  
$P_{\rm RSB}(q)$ drawn in Fig. 9c. 
Here $P(q; t_{\rm a}\simeq t_{\rm erg}^{\rm E})$ 
and $P_{\rm RSB}(q)$ (also those in Figs. 9a,b) are normalized as 
$\int_{-1}^1 P(q){\rm d}q = 2$, differently 
from those in Fig. 3.  The reason of this normalization is that 
time evolution of the clones at $t\sim$\tergE ($\ll$\tergL) 
of present interest is almost confined in a part of phase space 
with a common time-reversal symmetry.
[A small weight of $P(Q;t,t_{\rm w})$ smearing out to $Q<0$ is 
considered due to the finite-size effect]. 
It is rather surprising that near coincidence, 
$P(Q;t\sim t_{\rm erg}^{\rm E},t_{\rm w})$ 
$\simeq$ $P(q; t_{\rm a}\simeq t_{\rm erg}^{\rm E}(N, T))$ 
$\simeq$ $P_{\rm RSB}(q)$, holds for $Q,q\nge0.3$. 
[To be more accurate, near equality 
$P(q; t_{\rm a})$ $\cong$ $P_{\rm RSB}(q)$ for all $q$ 
is obtained only at 
$t$ larger than \tergL\ as seen in Fig. 3 and in the previous 
work\cite{BhattY}. By `near coincidence' here we disregard
quantitative difference in the weights associated with the peak at 
$Q,q \simeq$\qEA\ and qualitative difference in the shapes at 
$Q,q \nle 0.3$.]

The near coincidence $P(Q;t\sim t_{\rm erg}^{\rm E},t_{\rm w})$ 
$\simeq$ $P_{\rm RSB}(q)$ alone indicates that 
the organization of states in one `pure state' centered 
at the lowest energy state ($S_\infty$ in Fig. 8) is 
similar to that of the whole phase space with a common time-reversal 
symmetry which gives rise to $P_{\rm RSB}(q)$.
The other near coincidence $P(Q;t\sim t_{\rm erg}^{\rm E},t_{\rm w})$ 
$\simeq$ $P(q; t_{\rm a}\simeq t_{\rm erg}^{\rm E})$
tells more since its r.h.s. quantity is the overlap between two 
configurations which start from random 
initial configurations independently chosen. 
Its plausible interpretation we can think of is that 
a basin of attraction of one `pure state' reached by our 
simulation in fact covers almost an entire phase space with a common 
time-reversal symmetry. 
This is what we have called the dynamic dominant pure-states 
picture in section 1. We regard the near coincidence of the three distribution 
functions as an evidence that the aging processes we have simulated
are stochastic dynamics of the system in the phase space, whose 
organization of states in equilibrium is described by 
the static dominant pure-states picture\cite{MPSTV}.
In other words, our present simulations yield, for the first time to
our knowledge, an evidence for the latter picture from a dynamical
point of view.

\vspace*{-3mm} 
\section{Conclusion}

We have proposed the scenario that aging processes at a fixed 
$T$ (and $h$) in the SK model, which are observed in the 
aging range of \tqe\nle$t$\nle\tergE, are growth 
processes of quasi-equilibrium domains, or 
stochastic dynamics of the system looking for 
dominant pure states which exhaust the probability weights 
in equilibrium. It qualitatively explains most of our simulated 
results. In our arguments extensive energies of states (solutions 
of the TAP equation), which are represented by 
the ordinate of Fig. 8, and which govern relative Boltzmann 
weights in each quasi-equilibrium domain, play a central role. 
If, however, we want to know more details of aging phenomena, 
such as an explicit functional form of \Cttw, we have to analyze 
the organization of the states in the direction of the abscissa of 
Fig. 8. For this purpose the difference in the extensive energies 
may not be so important than the above argument 
(imagine that in the SK model with $N\rightarrow
\infty$ the system is somewhere in the middle of the `funnel' of 
Fig. 8 at a finite time). The research in this direction is now going on.

Acknowledgements:
We are specially indebted to R. Orbach and Y.G. Joh since 
this work was initiated from discussions with them on their 
experimental findings. We are grateful to J.-P. Bouchaud, 
L.F. Cugliandolo and J. Kurchan for many stimulating discussions. 
We also acknowledge useful discussions with J. Hammann, 
K. Nemoto, P. Nordblad, M. Ocio, E. Vincent and A.P. Young. 
The computation in this work has been done using FACOM VPP500 
of the Supercomputer Center, Institute for Solid State Physics, 
University of Tokyo. 
One of the authors (HY) was supported by Fellowships of the Japan
Society 
for the Promotion of Science for Japanese Junior Scientists.
This work is supported by a Grant-in-Aid for 
International Scientific Research, 
``Aging Phenomena in Complex Systems'' 
(\#08044060), and by a Grant-in-Aid for 
Scientific Research (\#08640477), 
both from the Ministry of Education, Science and Culture, Japan

\newpage
\bibliography{bibcluster}

\newpage

\noindent
{\bf Figure Captions}

\begin{enumerate}
\renewcommand{\labelenumi}{Fig.\arabic{enumi}:}

\item Time evolution of energy at $T=0.4$. a) Energy per spin in systems 
with $N(N_{\rm s})=$32(2000), 64(2000), 96(2000), 128(1000), 
192(600), 256(500), 512(200), 1024(40), and 2048(50) from top to 
bottom. b) Extensive energy relative to $E_{\rm eq}$ for 
$N=32\sim512$. The data with broken line represents 
$\Delta E(t_{\rm a}) + E_{\rm B}(t_{\rm a})$ for $N=512$ (see text for 
definition of $E_{\rm B}$).

\item The system-size ($N$) dependence of characteristic relaxation 
times related to aging processes in the SK model at $T=0.4$ for 
$N \le 256$; 
\tergL\ and \tergS\ of eq.(2) and \tergE\ determined by 
$\Delta E(t_{\rm erg}^{\rm E})$ $=0.5\ (=1.2T)$ from Fig. 2b. 
The solid line is ln(\tergS)$=2.56N^{1/4}-0.66$ due to Ref.\cite{MacYoun}. 
The area below \tergE\ is the aging range introduced 
and studied in the present work. 

\item Parisi's overlap distribution function $P(q; t_{\rm a})$ 
for $N=128$, $T=0.4$. It is normalized as 
$\int_{-1}^1 P(q; t){\rm d}q = 1$. The data are $t_{\rm a}=$16, 
64, 512, 4096 and 32768 from top to bottom at around $q\simeq0$.
The thick broken line represents $P_{\rm RSB}(q)$.

\item Auto-correlation functions \Cttw\ with \tw$=2^n$ ($n=3\sim13$) 
plotted against $t$ (a) and $t/$\tw\ (b) \ ($T=$0.4, $N=$512, 
$N_{\rm s}=$200). 

\item Auto-correlation functions $C(t_{\rm a}, t_{\rm a}/2)$ for 
some $T$ and $N$. 

\item Clones-correlation functions \Qttw\ with \tw$=$8, 16, 64, 256 
and 1024 ($T=$0.4, $N=$512, $N_{\rm s}=$100). Auto-correlation 
functions \Cttw\ are also shown by lines with the open symbols. 
The arrow indicates $t=$\tergE.

\item Induced magnetization \mttw\ (symbols with solid lines) 
and \mCttw\ (broken lines) of eq.(4) with \tw$=2^n$ ($n=3\sim13$) 
plotted against $t/$\tw\ ($T=$0.4, $h=$0.1, $N=$512, 
$N_{\rm s}=$200).

\item Schematic representation of the rugged energy structure 
of one `pure state' of the SK model. 
The dots represent local energy minima
 ($S_1, S_2,$ $\cdots,$ $S_{\infty}$) and the shaded areas 
quasi-equilibrium domains ($R_1, R_2, \cdots$).

\item a) Distribution functions of clones-correlation $P(Q;t,t_{\rm
    w})$ corresponding to the curve with \tw=64\ in Fig. 6. The broken 
  curve without symbols represents $P_{\rm RSB}(q)$ (also in
  Figs. 9b,c, and 10) 
b) Comparison of $P(Q;t,t_{\rm w})$ with $N=128, 512$ and \tw=64: the
solid curve 
with circles represents $P(Q;t=512\simeq t_{\rm erg}^{\rm E},t_{\rm w})$ 
of $N=128$, the broken curve with triangles $P(Q;t=2^{15}\simeq t_{\rm
  erg}^{\rm E},t_{\rm w})$ of $N=512$, and the thin solid curve 
$P(Q;t=2^{15}\simeq t_{\rm erg}^{\rm L},t_{\rm w})$ 
of $N=128$. 
c) Comparison of $P(Q;t\sim t_{\rm erg}^{\rm E},t_{\rm w})$ 
(broken curve with squares), $P(q; t_{\rm a}\simeq t_{\rm erg}^{\rm E})$ 
(solid curve), and $\simeq$ $P_{\rm RSB}(q)$. for
$N=512$ and $T=0.4$. In this figure $P(q)$  are normalized as 
$\int_{-1}^1 P(q){\rm d}q = 2$.

\item Distribution functions of auto-correlation $P(C;t,t_{\rm
    w})$ corresponding to the curve with \tw=64\ in Fig. 6.
\end{enumerate}

\nopagebreak

\pagestyle{empty}
\noindent{\large FIGURES}

\vspace*{5mm}
\hbox to \textwidth{
\vtop{
\hsize=15cm
\centerline{
        \epsfxsize=10.0cm
        \epsfysize=8.0cm
        \epsfbox{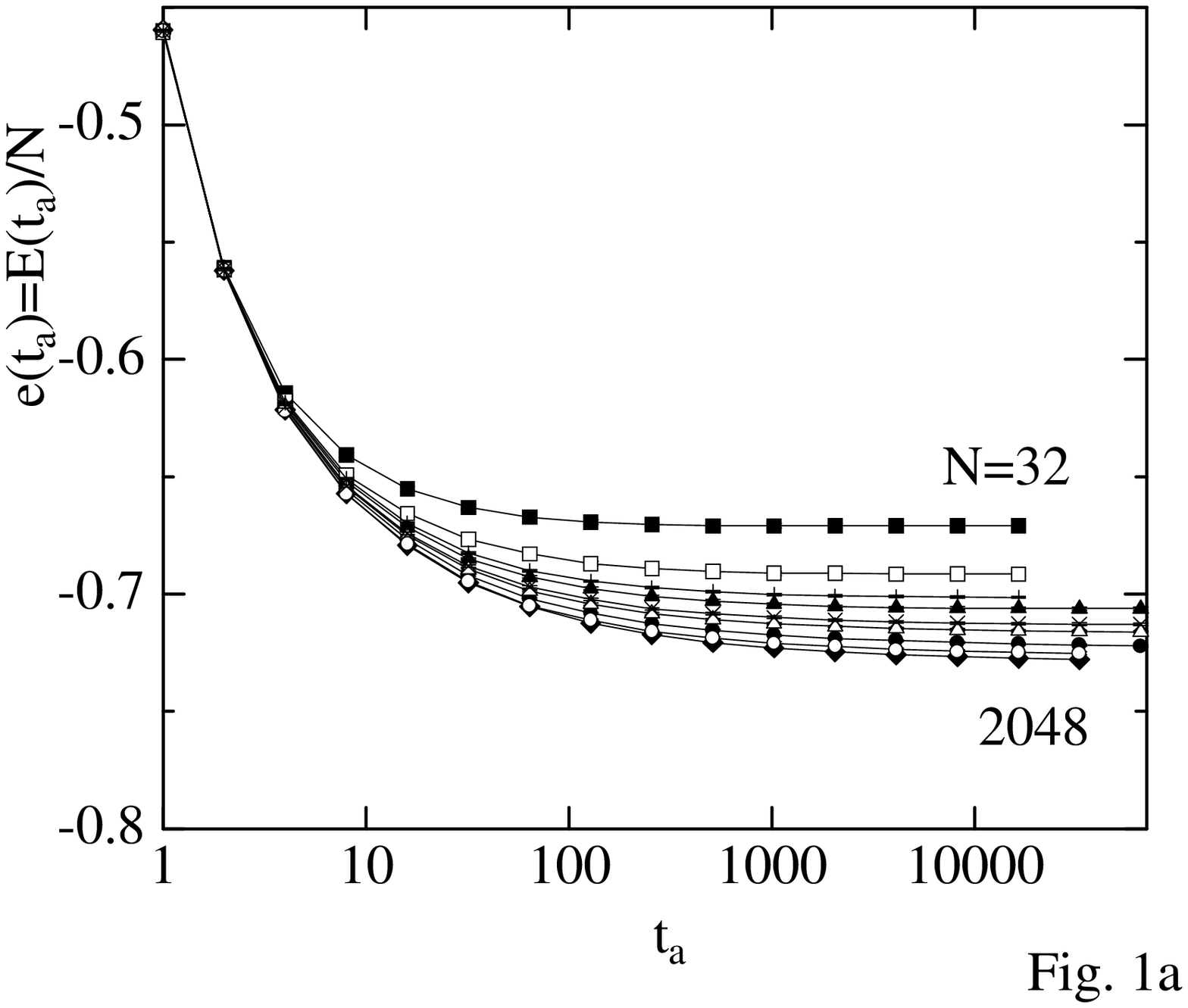}
}
\noindent
}}

\vspace*{5mm}
\hbox to \textwidth{
\vtop{
\hsize=15cm
\centerline{
        \epsfxsize=10cm
        \epsfysize=8cm
        \epsfbox{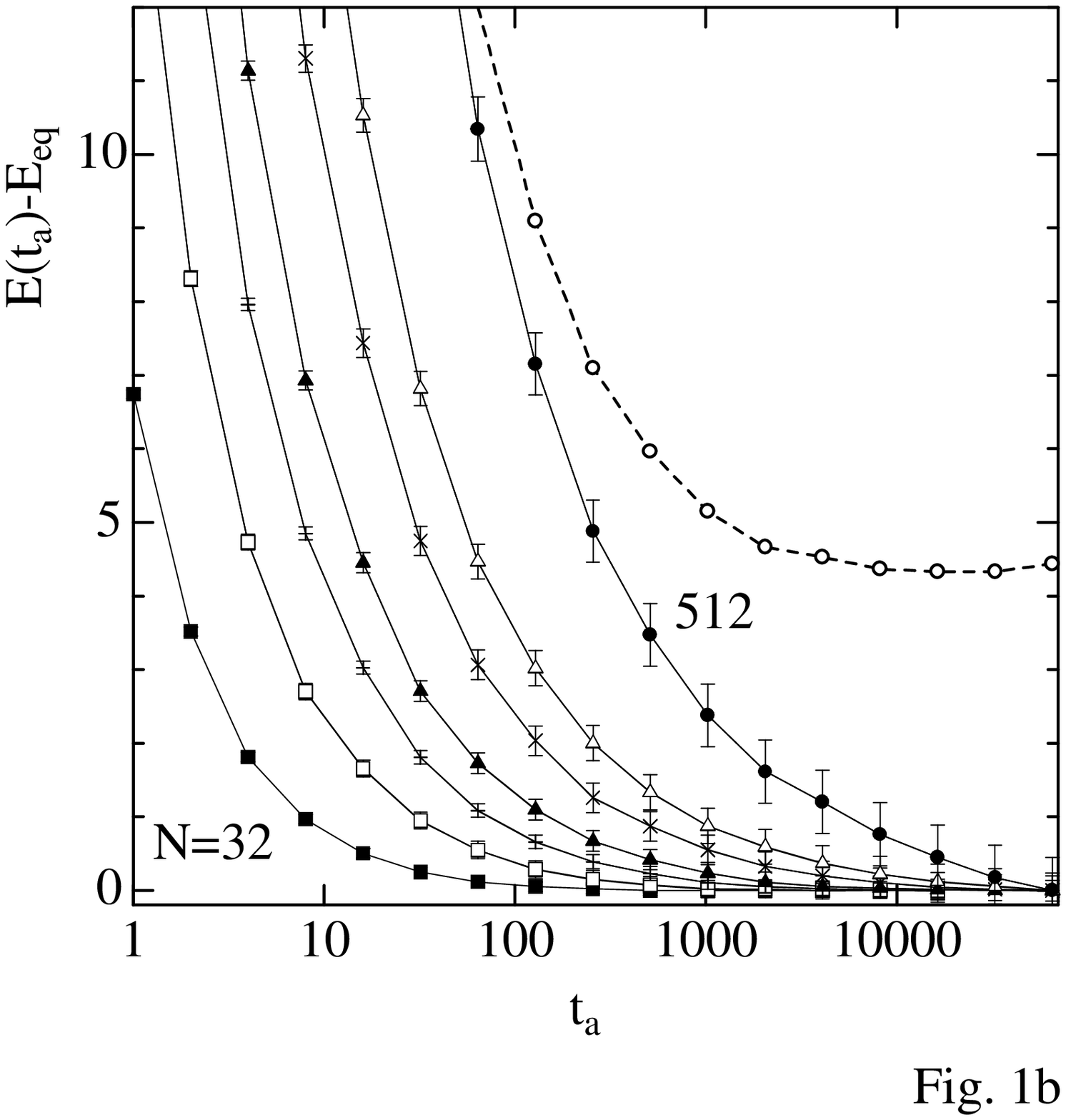}
}
}}

\vspace*{5mm}
\hbox to \textwidth{
\vtop{
\hsize=15cm
\centerline{
        \epsfxsize=10cm
        \epsfysize=8cm
        \epsfbox{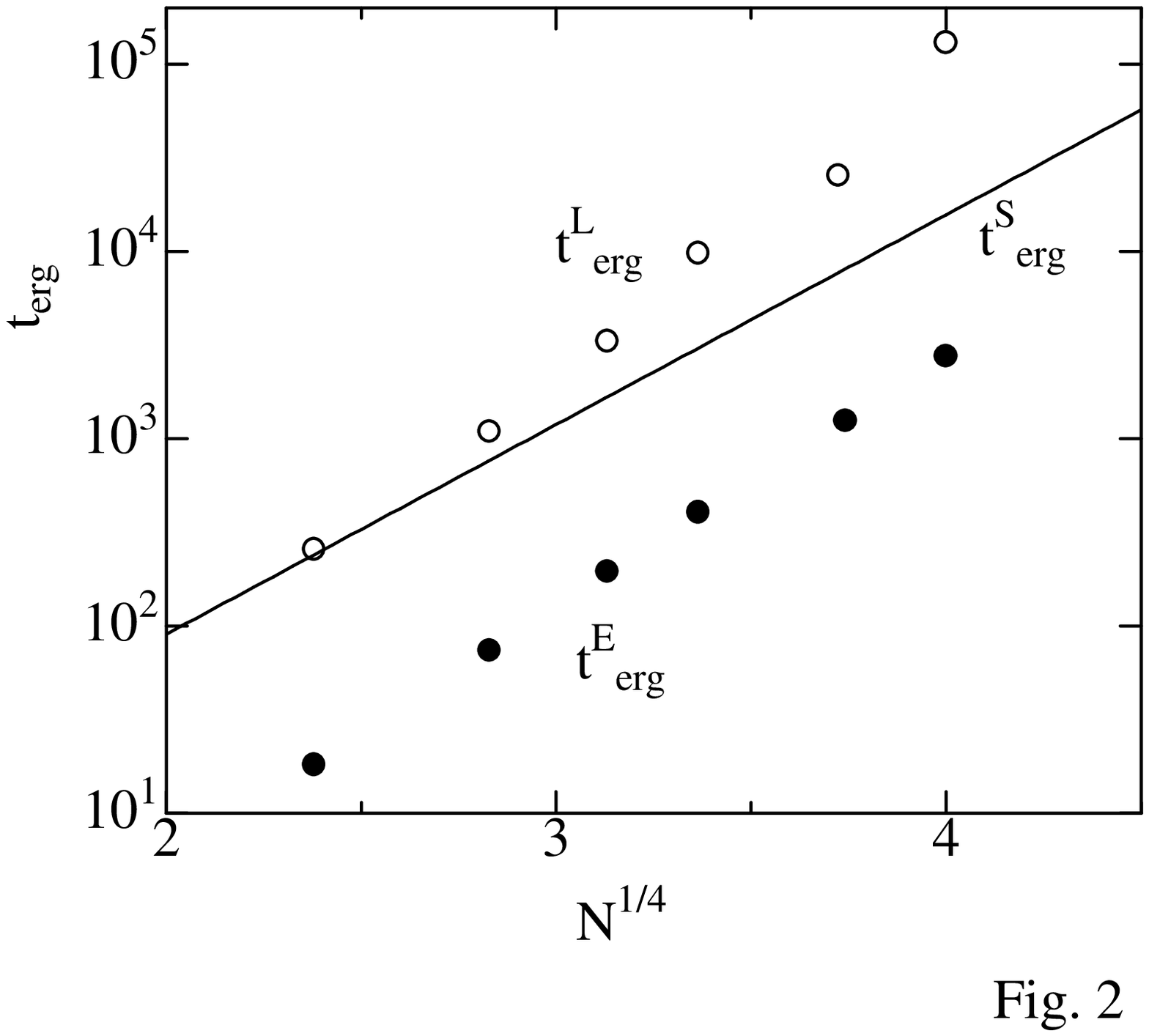}
}
}}

\newpage 
\vspace*{5mm}
\hbox to \textwidth{
\vtop{
\hsize=15cm
\centerline{
        \epsfxsize=10cm
        \epsfysize=8cm
        \epsfbox{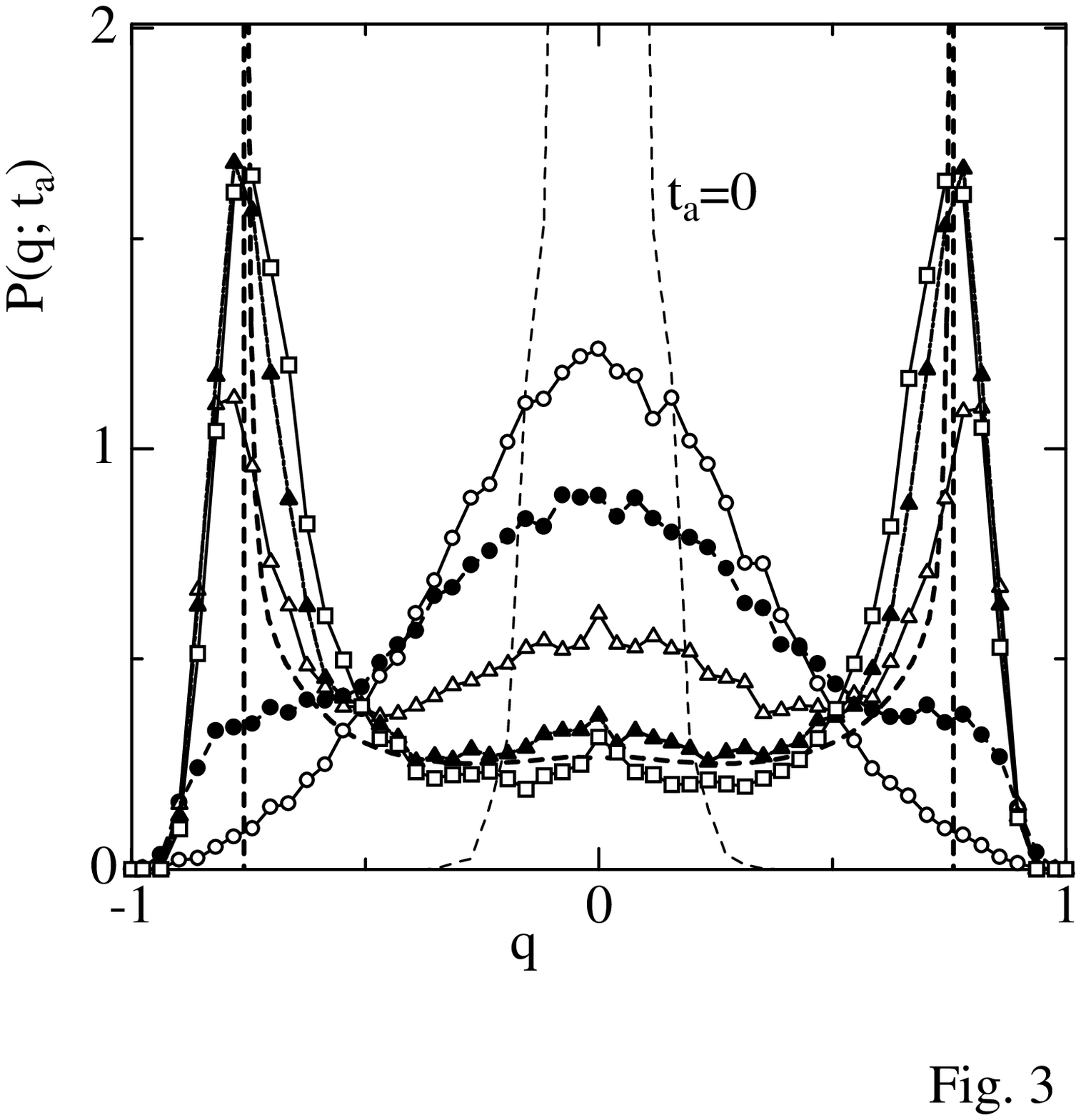}
}
}}

\vspace*{5mm}
\hbox to \textwidth{
\vtop{
\hsize=15cm
\centerline{
        \epsfxsize=10cm
        \epsfysize=8cm
        \epsfbox{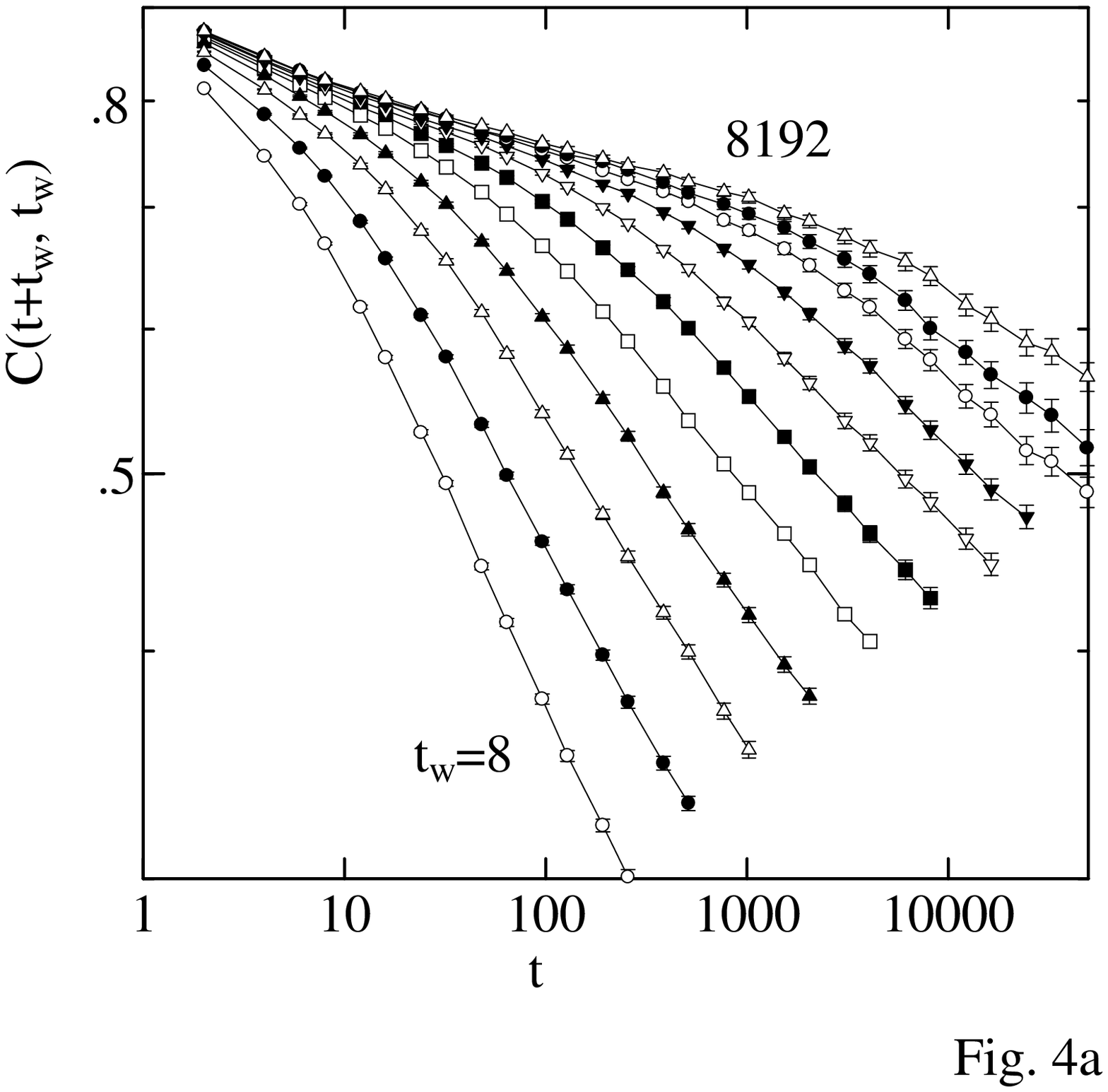}
}
}}

\vspace*{5mm}
\hbox to \textwidth{
\vtop{
\hsize=15cm
\centerline{
        \epsfxsize=10cm
        \epsfysize=8cm
        \epsfbox{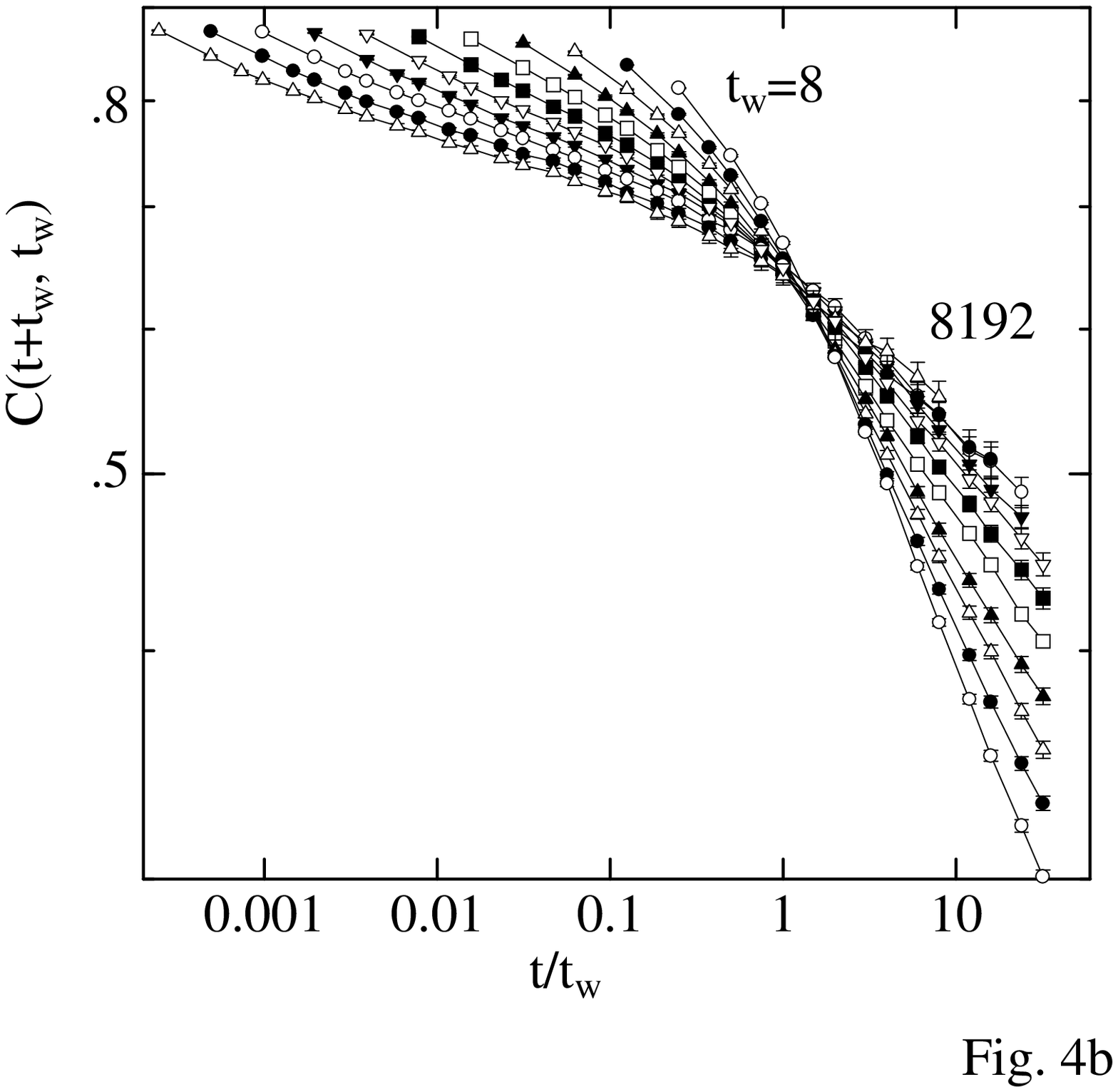}
}
}}

\newpage
\vspace*{5mm}
\hbox to \textwidth{
\vtop{
\hsize=15cm
\centerline{
        \epsfxsize=10cm
        \epsfysize=8cm
        \epsfbox{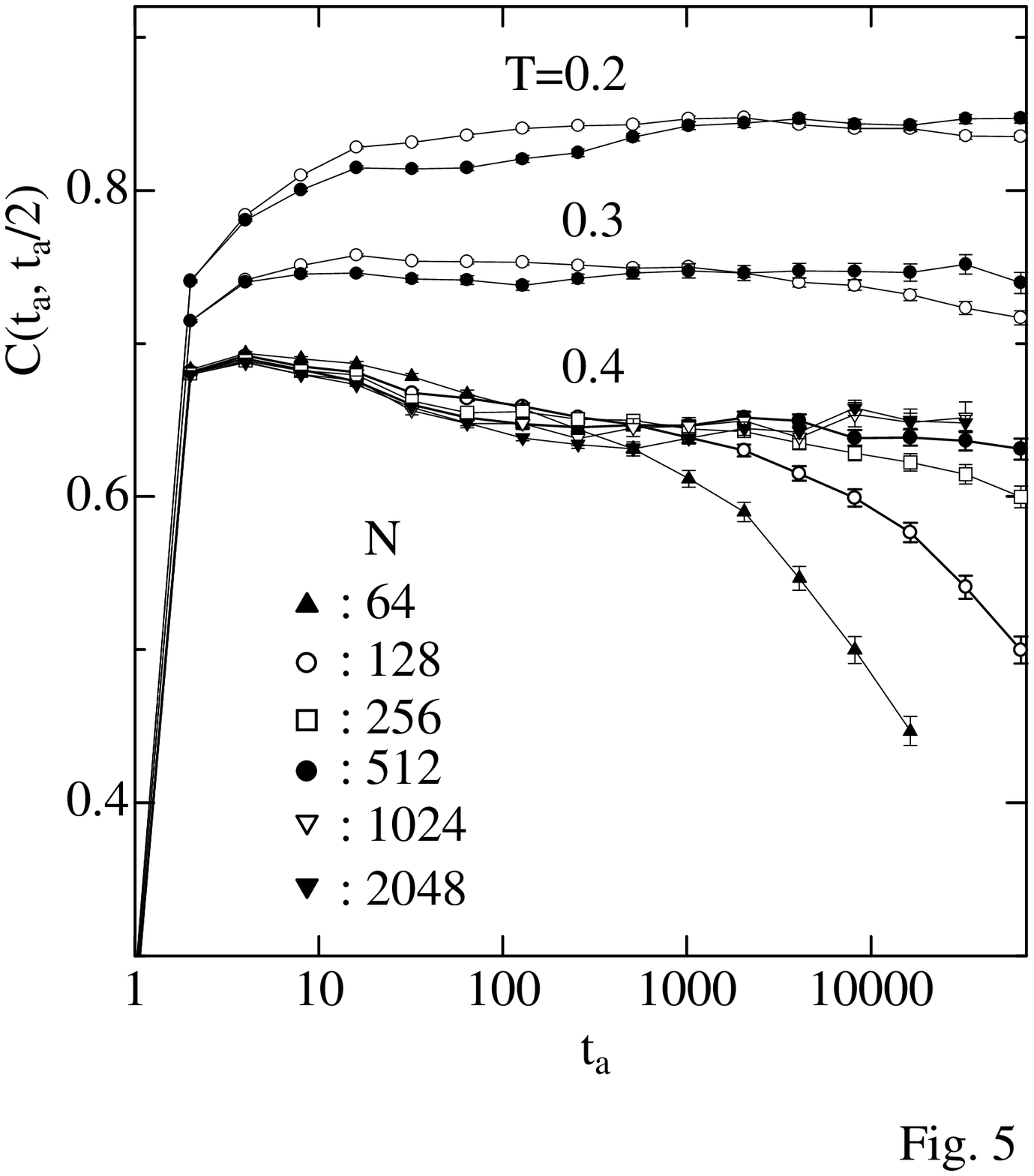}
}
}}

\vspace*{5mm}
\hbox to \textwidth{
\vtop{
\hsize=15cm
\centerline{
        \epsfxsize=10cm
        \epsfysize=8cm
        \epsfbox{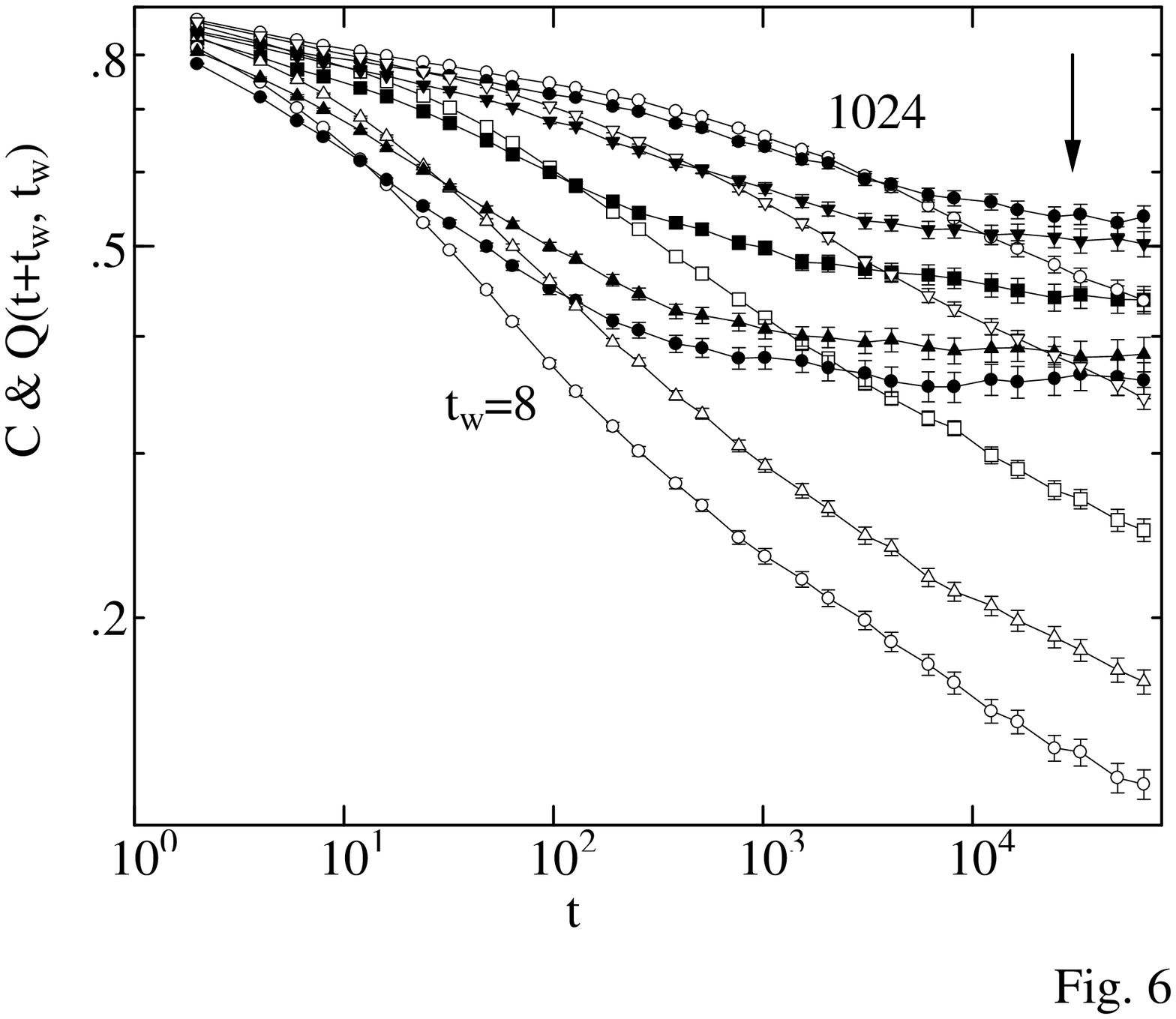}
}
}}

\vspace*{5mm}
\hbox to \textwidth{
\vtop{
\hsize=15cm
\centerline{
        \epsfxsize=10cm
        \epsfysize=8cm
        \epsfbox{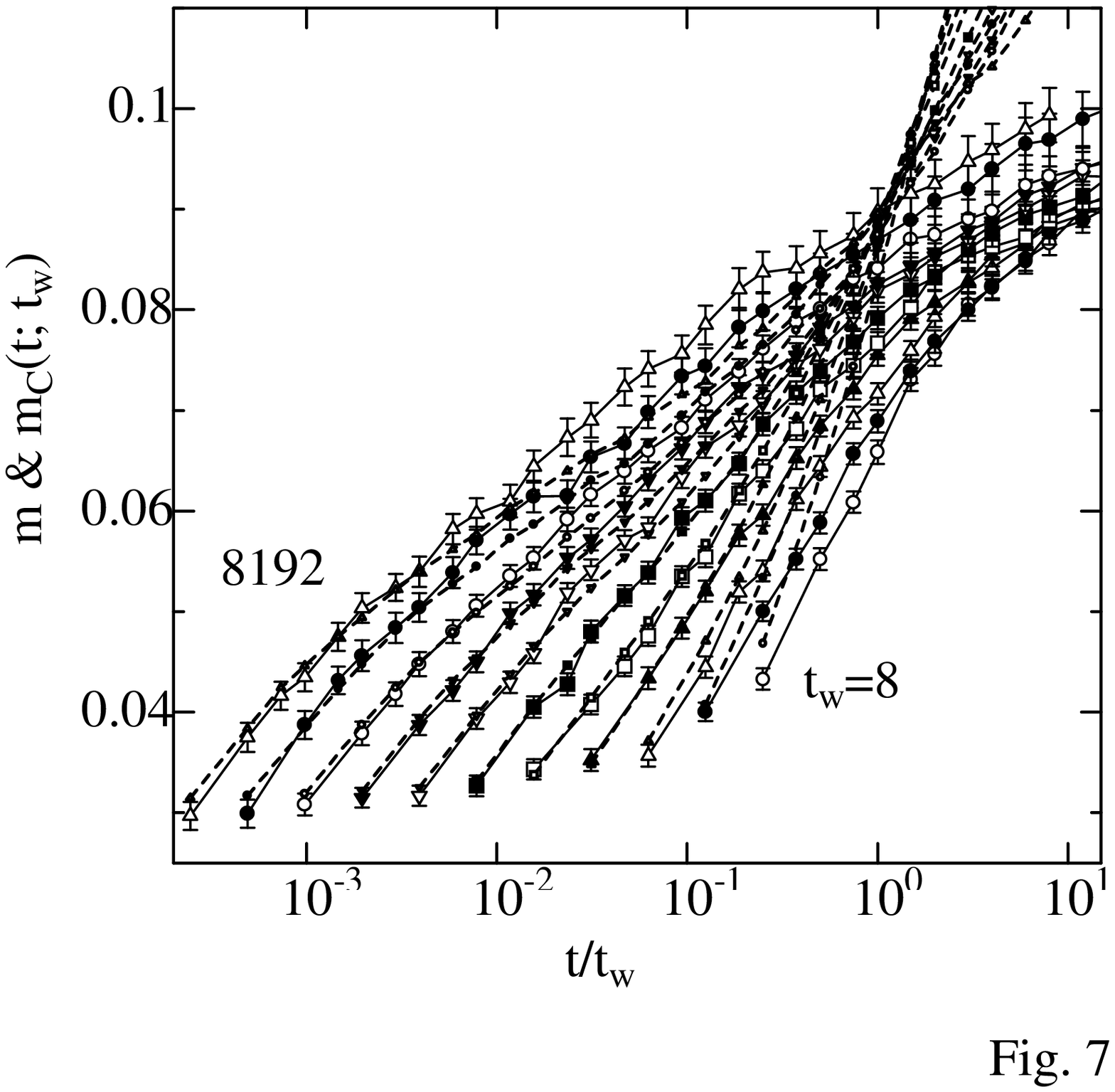}
}
}}

\newpage
\vspace*{5mm}
\hbox to \textwidth{
\vtop{
\hsize=15cm
\centerline{
        \epsfxsize=10cm
        \epsfysize=8cm
        \epsfbox{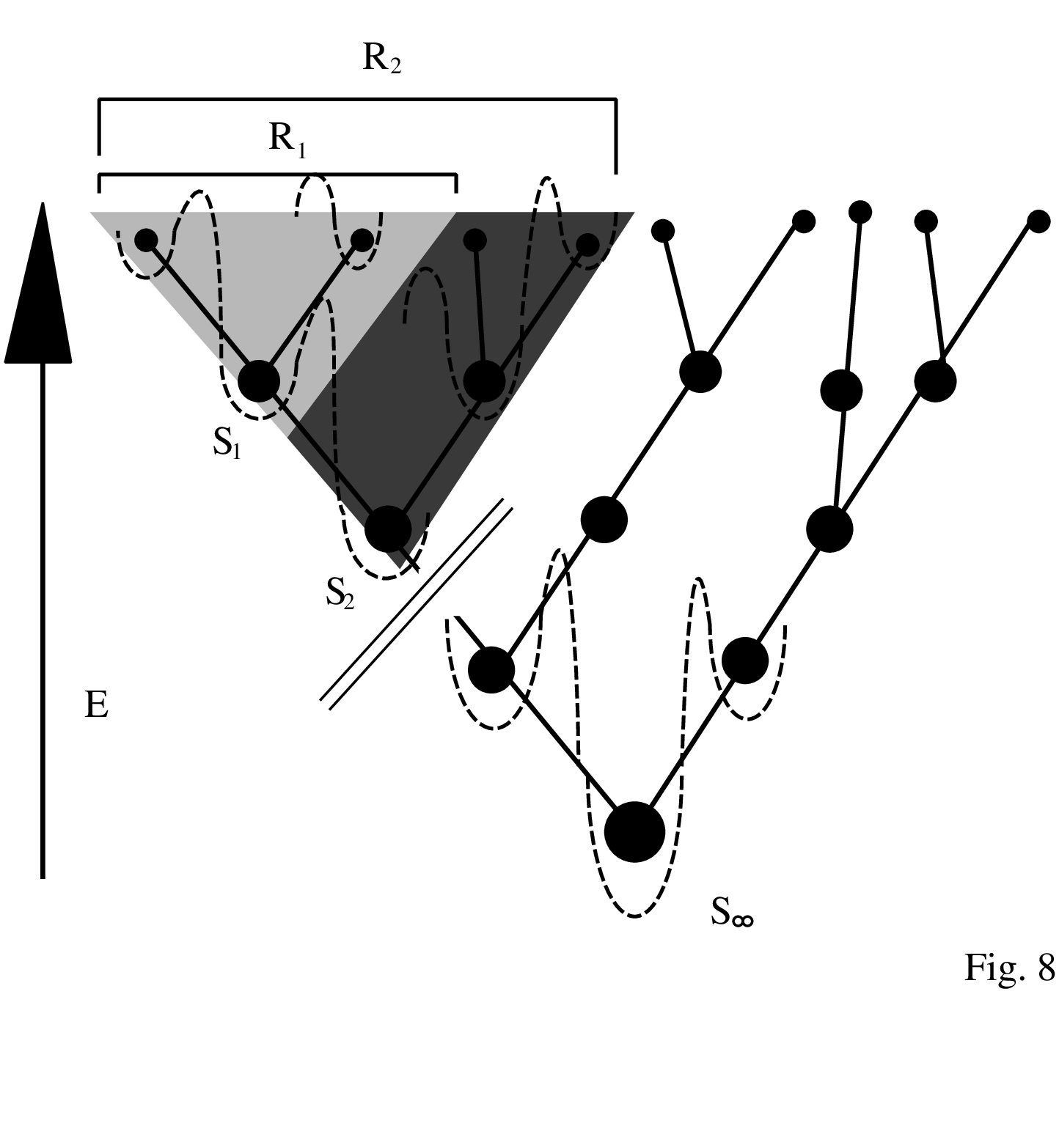}
}
}}

\vspace*{5mm}
\hbox to \textwidth{
\vtop{
\hsize=15cm
\centerline{
        \epsfxsize=10cm
        \epsfysize=8cm
        \epsfbox{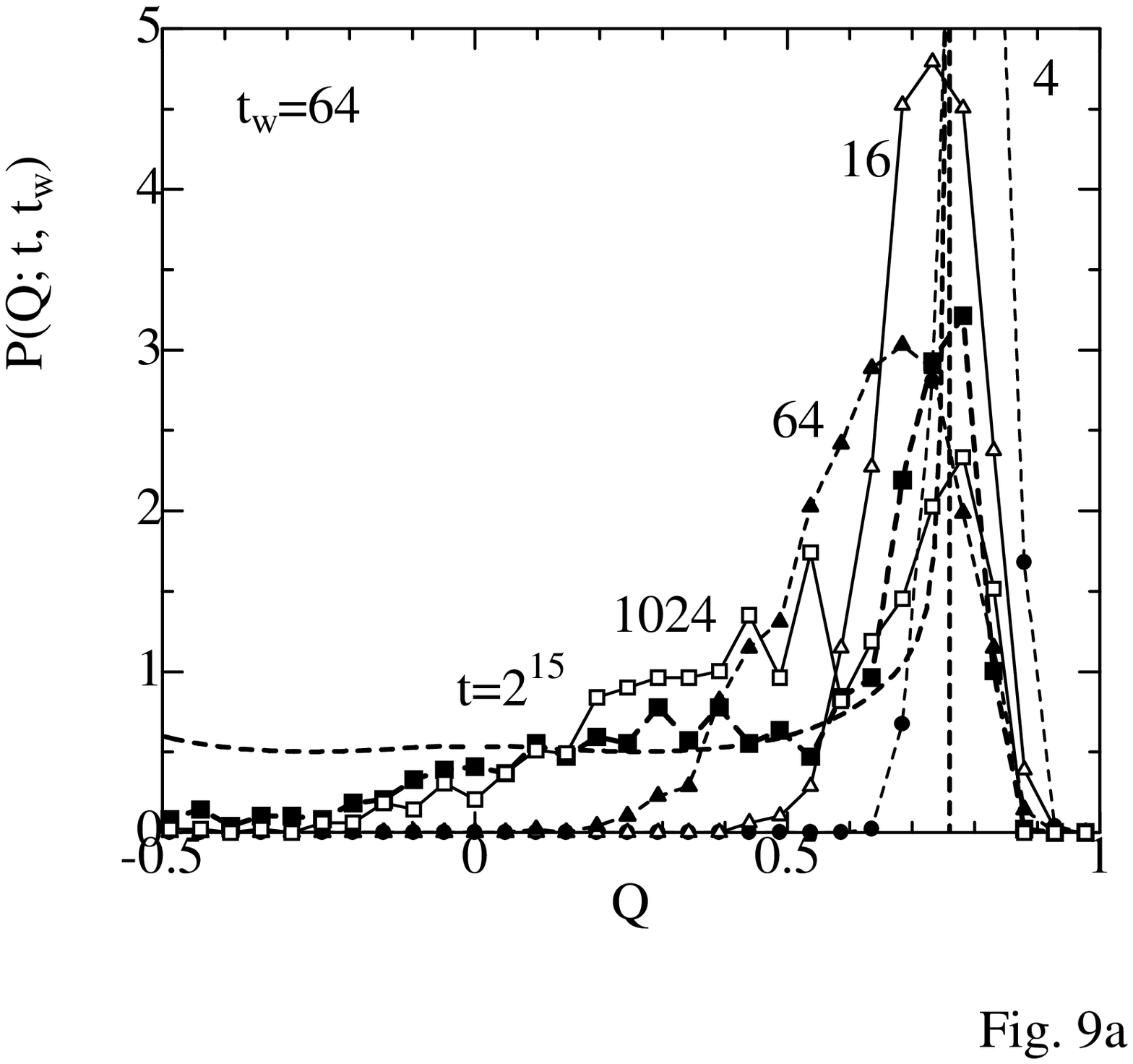}
}
}}

\vspace*{5mm}
\hbox to \textwidth{
\vtop{
\hsize=15cm
\centerline{
        \epsfxsize=10cm
        \epsfysize=8cm
        \epsfbox{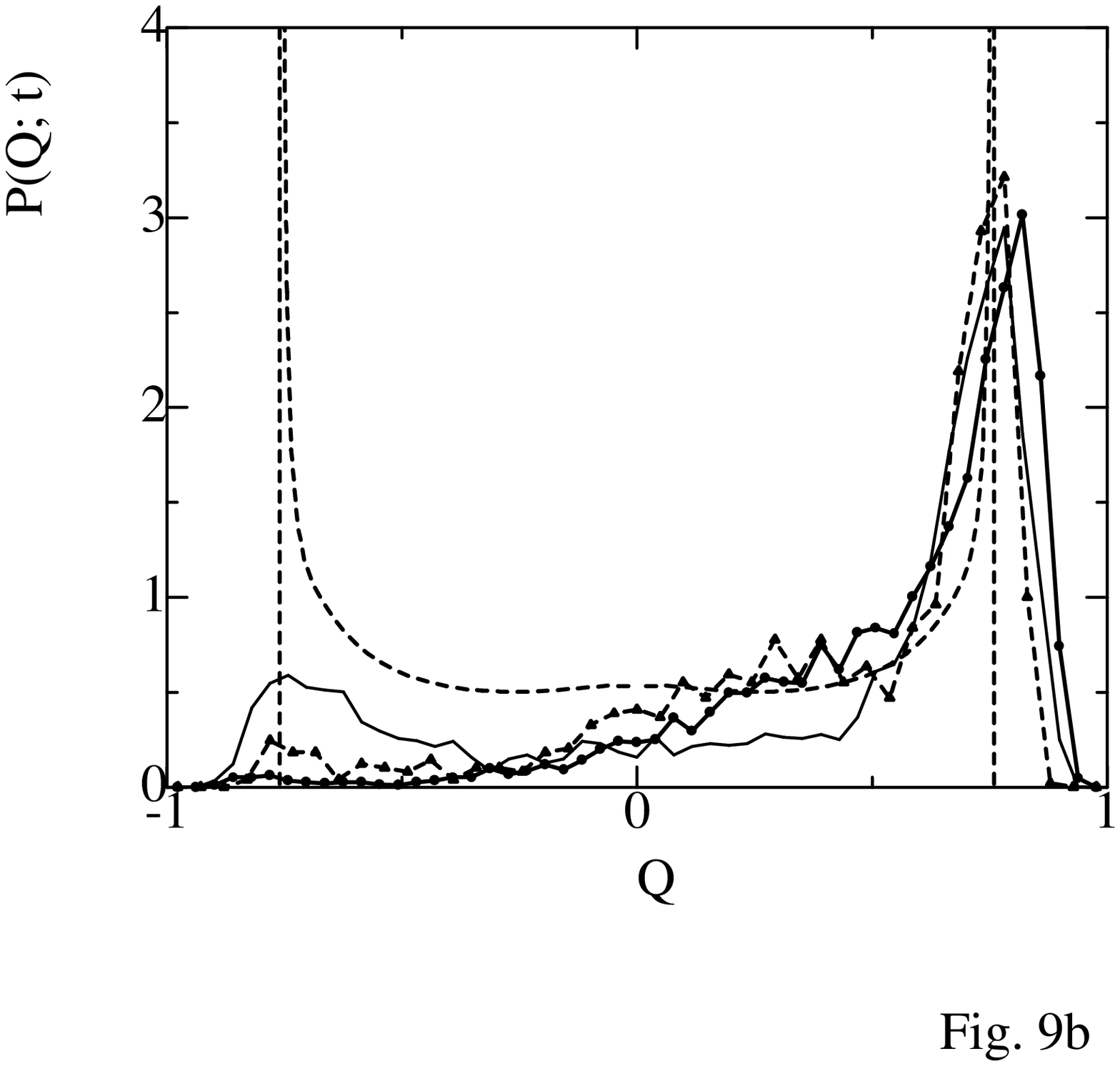}
}
}}

\newpage
\vspace*{5mm}
\hbox to \textwidth{
\vtop{
\hsize=15cm
\centerline{
        \epsfxsize=10cm
        \epsfysize=8cm
        \epsfbox{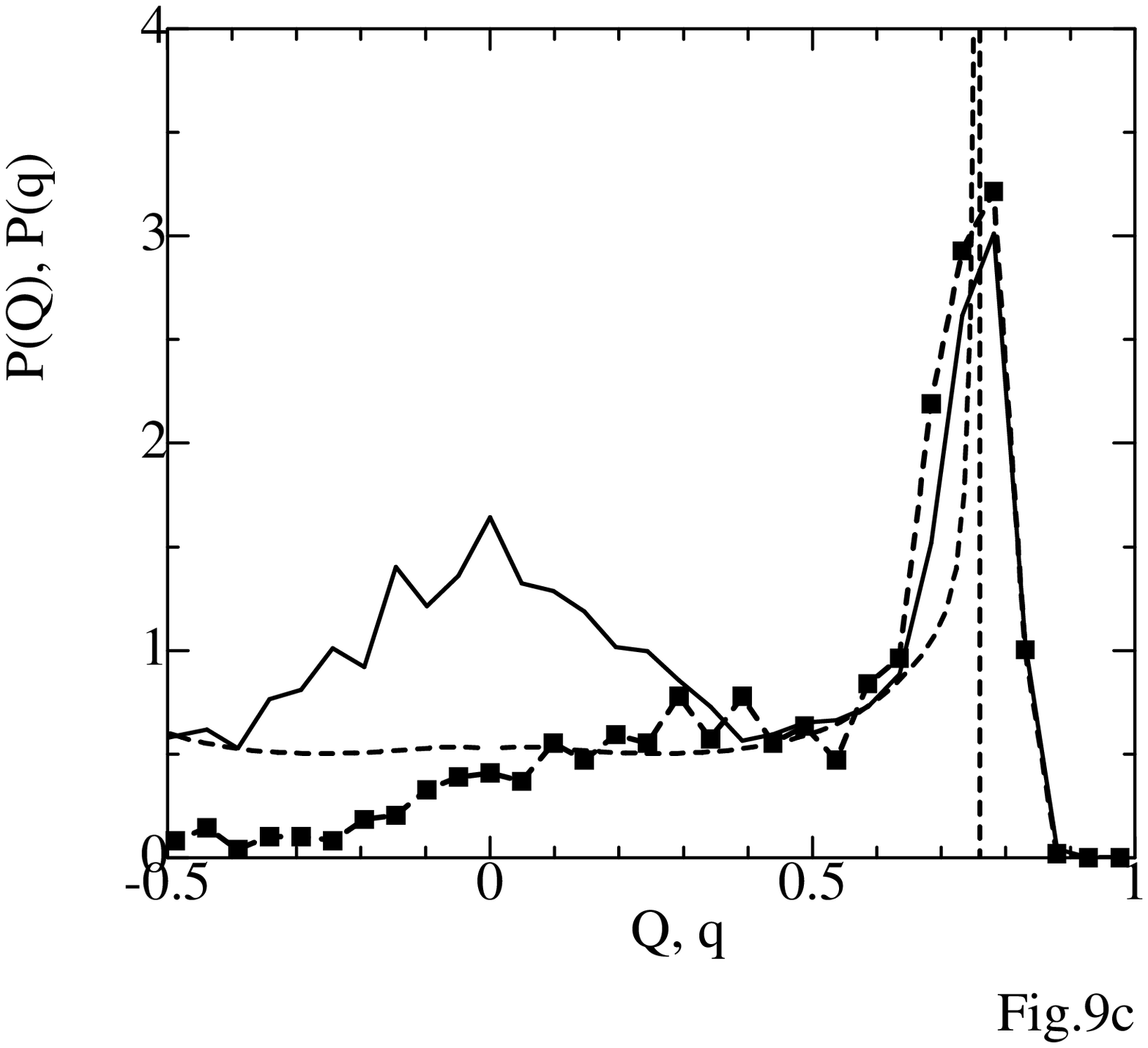}
}
}}

\vspace*{5mm}
\hbox to \textwidth{
\vtop{
\hsize=15cm
\centerline{
        \epsfxsize=10cm
        \epsfysize=8cm
        \epsfbox{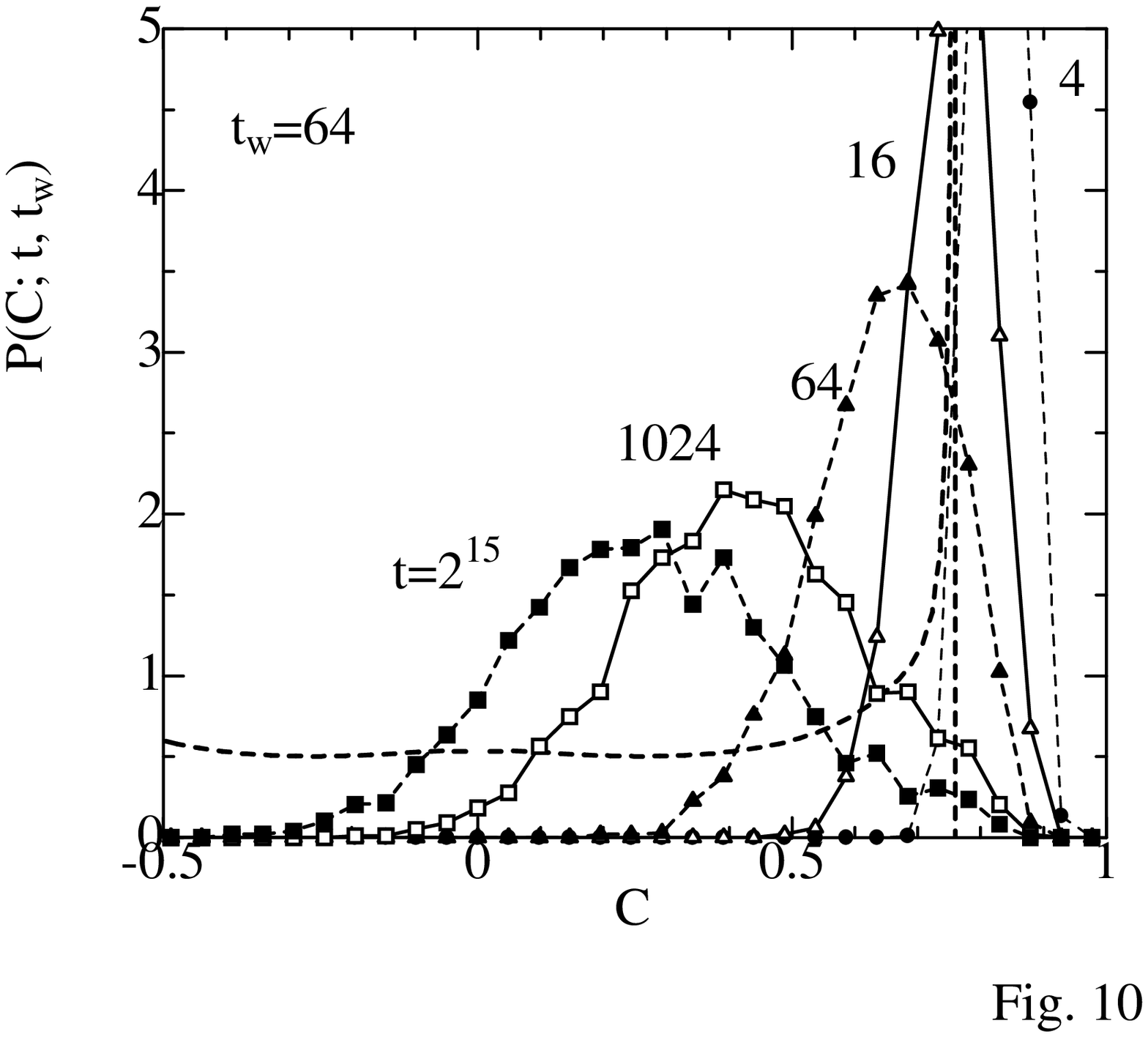}
}
}}

\end{document}